# A Tri-Mode Coupled Coil with Tunable Focal Point Adjustment for Bio-Medical Applications

Raunaq Pradhan, *Student Member, IEEE,* Xiaohua Feng, and Yuanjin Zheng*, *Member, IEEE*

*Abstract*— *Objectives:* The paper proposes the design of a tri-mode coupled coil which enables three modes of operation for inducing electromagnetic field, where the focal adjustment of the E field can be optimized. Methods: The setup consists of two identical figure-of-eight coils, namely coil 1 and coil 2, coupled to each other by magnetic resonance coupling. Coil 1 is driven by active source, where coil 2 is driven by the magnetic field coupled from coil 1. The frequency of operation would affect the coupling between the coils and hence the current ratios induced. Results: In the first and the second modes, the current dominates at the first coil and the second coil, respectively. In the third mode, both coils conduct similar amount of currents. Conclusion: The concept is proven by measuring the current ratios of the coils and the voltage induced in biological tissues. The current ratios in the first and the second modes are measured as maximum 13.0 and minimal 0.258 at frequencies 487.7 kHz and 453.2 kHz, respectively, while the current ratio measured in the third mode is 1.03. Significance: The tri-mode coil could potentially be applied for biological applications such as pulsed electromagnetic energy treatment and thermoacoustic imaging.

## NOMENCLATURE

**E** = Electric field intensity (V/m)
**B** = Magnitude flux density (T)
**J** = Current density (A/m$^2$)
$f_0$ = Self-resonance frequency (Hz)
$f_m$ = Lowest resonance frequency (Hz)
$f_e$ = Highest resonance frequency (Hz)
$f_1$ = First operation frequency (Hz)
$f_2$ = Second operation frequency (Hz)
$f_3$ = Third operation frequency (Hz)
$\mu_0$ = Permeability of the free space (H/m)
$\sigma$ = Electrical conductivity (S/m)
$N$ = Number of turn of the coil
$I$ = Magnitude of current flowing through the coil (A)
$I_1$ = Magnitude of current flowing through coil 1 (A)
$I_2$ = Magnitude of current flowing through coil 2 (A)
**dl**' = Differential coil element
**r**' = Position vector of the differential element **dl**'
**r** = Position vector of the observation point
$x, y, z$ = Cartesian coordinates

$D$ = Distance between coils 1 and 2 (m)
$L_m$ = Mutual inductance (H)
$L$ = Self-inductance (H)
$C$ = Capacitor cascaded in series to the coil for resonance
$R$ = Coil resistance plus the resistance of the small resistor cascaded in series to the coil (Ω)
$L_1, L_2$ = Inductor values in the multi-impedance matching network (H)
$C_1, C_2, C_3$ = Capacitor values in the multi-impedance matching network (F)
$R_S$ = Source resistance (Ω)
$V_S$ = Source voltage (V)
$G_{I1}$ = Transconductance between the coil currents $I_1$ and the source voltage $V_S$ (A/V)
$G_{I2}$ = Transconductance between the coil currents $I_2$ and the source voltage $V_S$ (A/V)
$G_{E1}$ = Electric field gain between the electric field $E_1$ induced at the observation point (contributed by the coil current $I_1$) and the product of the coil current $I_1$ and frequency (V/AmHz$^{-1}$)
$G_{E2}$ = Electric field gain between the electric field $E_2$ induced at the observation point (contributed by the coil current $I_2$) and the product of the coil current $I_2$ and frequency (V/AmHz$^{-1}$)
$\theta$ = Phase difference between the current across the two coils 1 and 2 (°)
$d$ = The distance between the two terminals of the voltage measurement (m)
$P_\sigma$ = Ohmic power density (W)

## I. INTRODUCTION

In biomedical engineering, magnetic coils [1] are often used to induce electromagnetic field and hence electromotive force to conduct current in the biological mediums. It replaces the traditional approaches that applies electrodes on the skin directly to conduct current [2] which is usually due to lack of deep penetration and localization. This is because more current is expected to flow near the surface of the skin instead, where the path is shorter resulting in lesser resistance. Using magnetic coils to induce current is also better compared to some other approaches which involve implantation of electrodes/ coils under the skin [3] – [5], in the sense of being non-invasive. The applications of magnetic coils include magnetic stimulation [6] – [8], pulsed electromagnetic field therapies [9] – [11], thermoacoustic imaging [12] – [14], and radio frequency (RF) ablation [15] – [17]. The current density **J** is proportional to the electric field **E** induced in the biological medium, where the proportional constant is the conductivity [18]:

$$\mathbf{J} = \sigma \mathbf{E} \qquad (1)$$

Hence, in this paper the magnitude of the **E** field generated by the coils will be the main consideration. On the other hand, the geometrical distribution of the **E** field is also a main concern. Programmable subwavelength coil arrays to generate focused electric field and magnetic field patterns (with the help of switches) for biomedical applications have been described earlier in literature [19] - [21]. Considering the case of magnetic stimulation, a sufficient large current should be induced at the target only, while the current induced in the other areas should be minimized. Therefore, it will be beneficial to design a coil where the distribution of the **E** field can be adjusted so as to optimize the current induced at different locations.

To summarize, since the **E** field governs the magnitude of current induced in the biological medium as given in equation (1), the magnitude of **E** field that can be generated by the coils and its distribution would be the primary interest or performance indicator of the coil design. The magnetic flux density **B** would be the secondary interest, due to its relationship to the **E** field by Faraday's law:

$$\nabla \times \mathbf{E} = -\frac{\partial \mathbf{B}}{\partial t}, \qquad (2)$$

This indicates that the varying **B** field would induce **E** field which is commonly known as electromagnetic induction. However, using **E** field for analysis would be much simpler as being a more direct indicator as stated in equation (1).

In this paper, a coil setup which can adjust the electric field distribution by changing the operation modes is designed. The block diagram of the setup is given in Fig. 1. It consists of two identical figure-of-eight coils, namely coil 1 and coil 2, coupled to each other by magnetic resonance coupling [22] – [24]. The detailed structure and operation principle of the system will be explained in the remaining sections. The proposed coil circuit setup supports three operation modes. The difference between the three modes is characterized by the ratio between the magnitude of currents conducting coils 1 and 2, which results in different **E** field distributions among the modes. The combination of the three modes could be applied to optimize the **E** field distribution based on the location of the excitation target for the sake of minimizing the **E** field out of the target.

The setup will be referred as tri-mode coupled coil in this paper, which is applied to optimize the electromagnetic field distribution for focal tuning. Furthermore, based on the tri-mode coil configuration, a formula based methodology is developed for tunable focal point adjustment within the two coils by selecting the operation mode and the input voltage. Although the coupled coils using the magnetic resonance coupling (MRC) [22] – [26] has been proposed, those designs are not for changing the distribution of **E** field that can adjust the stimulation location. Meanwhile, in this paper, the distribution of the **E** field is designed to be adjusted according to the location to be excited.

Although it is also possible by moving an actively driven external coil to change the **E** field instead of changing the operation mode, it is supposed that the mechanical movement by motors has a slower response time than the change of operation mode which can be electrically controlled. Moreover, the proposed device is steady during operation, while the device would be moving during operation if it depends on the mechanical approach. Last but not least, the proposed method applies one active source only to coil 1, which does not require another active source to drive coil 2. It could simplify the system and hence operating at a lower cost.

The paper is organized as follows: In Section II, the theory and the design of the tri-mode coupled coil will be introduced. In Section III, the electric field distribution of the tri-mode operation will be simulated. In Section IV, the procedures of tuning focus with the coils will be introduced. In Section V, the coil measurement result will be presented. In Section VI, an ex-vivo tissue experiment will be done to verify the tri-mode concept for the electromagnetic induction. In Section VII, the discussion on several practical issues will be included for further clarification of the design. Finally, the conclusion will be given in Section VIII.

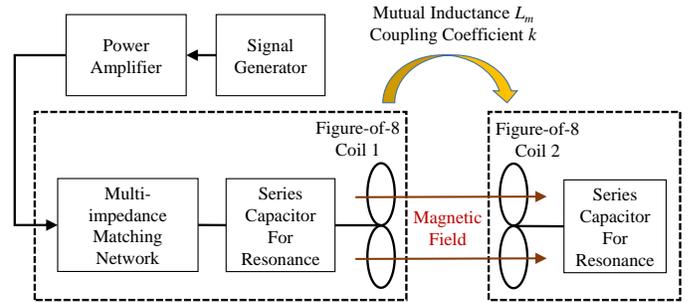

Fig. 1. Block diagram of the tri-mode couple coil setup.

## II. TRI-MODE COUPLED COIL DESIGN

In this section, the theories of the tri-mode coupled coil design with the multi-impedance matching network will be introduced.

### A. System and Coil Configuration

The block diagram of the setup is given in Fig. 1. It consists of two identical figure-of-eight coils separated by $D = 9$ cm, namely coil 1 and coil 2, coupled to each other by magnetic resonance coupling [22] – [24]. Coil 1 is driven by signal generator, which is cascaded to the power amplifier and the multi-impedance matching network. Coil 2 is driven by the magnetic field coupled from coil 1 only without any other source. Then, the 3D model of the coupled coils 1 and 2 is illustrated in Fig. 2, where each of them has 20 turns and coil diameter of 9 cm. The distance between the two coil centers in each of the figure-of-eight coil is 5 cm. The figure also shows the current direction respective to a time instant while the polarity alternates sinusoidally. The proposed coil circuit setup supports three operation modes at three adjacent frequencies which have different input impedance. Therefore, the multi-impedance matching network in Fig. 1 is used to match the coil so that the coil can be used for the three operation modes.

*B. Circuit Configuration*

Firstly, the pair of figure-of-eight coils with self-inductance $L$ (as illustrated in Fig. 1) is cascaded with capacitor $C$ as depicted in Fig. 3. A small resistor (0.11 Ω) is cascaded in series with the resonator so that the current flowing through the coil can be measured by detecting the voltage across the resistor. The cascaded reactive components are selected such that the resonant frequencies at the both coils are the same for magnetic resonance coupling. For demonstration purposes, the resonant frequency is chosen as 450 kHz which is at the mid-range of the pulsed electromagnetic energy therapy [9, Table 3]. Meanwhile, the resonating frequency and thus the application frequency are adjustable by changing the values of the capacitors which resonate with the coils. The self- and mutual-inductances of the coil are measured to be $L$ = 106 μH and $L_m$ = 2.51 μH, while the capacitor cascaded is $C$ = 1.14 nF. Hence, the input impedance of the circuit is given in Fig. 4. It can be noticed that there are three resonance frequencies where the reactance equals zero. The first one is the self-resonance frequency, given by [22]

$$f_0 = \frac{1}{2\pi\sqrt{LC}} \quad (3)$$

which results as $f_0$ = 457.7 kHz.

The additional second and the third resonance frequencies (lowest resonance frequency $f_m$ and highest resonance frequency $f_e$) are resulted from the mutual inductance $L_m$ between the two coils, given by:

$$f_m = \frac{1}{2\pi\sqrt{(L+L_m)C}}, \quad (4)$$

$$f_e = \frac{1}{2\pi\sqrt{(L-L_m)C}}, \quad (5)$$

which are calculated as $f_m$ = 452.6 kHz and $f_e$ = 462.9 kHz.

*C. Tri-Mode Operations*

The coil will be applied for three adjacent operating frequencies with different current ratio, which are used for three operation modes.

In the first mode, the current is dominated at coil 1 while minimized at coil 2, so that the **E** field near coil 1 is large. This mode 1 is used to minimize the electromagnetic field distribution near coil 2 where is not near the focal point located near coil 1. In the second mode, the current is dominated at coil 2 while minimized at coil 1, which is counter-mode to the first mode. This mode 2 is used to minimize the electromagnetic field distribution near coil 1 where is not near the focal point located near coil 2. In the third mode, the magnitudes of the currents conducting at the two coils are the same, both **E** field near the two coils is large, and also superposed and enlarged the **E** field at the center between the two coils. As using only one coil would require very large current to generate adequate **E** field magnitude at the center which is far away from the coils, the mode 3 is used to prevent overloading the current at the either coil and the **E** field magnitude near it by utilizing both coils to superpose a large **E** field at the center.

For the operation principle, define $I_1$ and $I_2$ as the currents conducting coil 1 and coil 2, respectively. Then, the current ratio between coil 1 and coil 2, $I_1/I_2$ is given in Fig. 5.

At the first mode operation frequency $f_1$, $I_1>I_2$, the current is dominated at coil 1 whereas minimized at coil 2. The current ratio is selected to be large while not too far away (at around 5 % of $f_0$) from the resonance frequencies so that the three operation frequencies will be similar. Therefore, $f_1$ = 480 kHz is selected.

At the second mode operation frequency $f_2$, $I_2>I_1$, the current is dominated at coil 2 whereas minimized at coil 1. Therefore, the minimum point on the curve is chosen, which is the self-resonance frequency $f_2 = f_0$ = 457.7 kHz.

Although the first and the second operations are similar, they are not reductant because they are corresponding to the target at different location. Although this can also be achieved by switching the positions of the coils (with the active source connected to coil 1), the response time of the mechanical switching approach would be more significant as compared to the proposed electrical switching approach. Moreover, it would become a moving device instead of the proposed steady device.

At the third mode operation frequency $f_3$, $I_1=I_2$, the magnitudes of the currents conducting at coil 1 and coil 2 are the same. Therefore, the current ratio of 1 is selected, which is either $f_m$ or $f_e$. However, $f_3 = f_m$ = 452.6 kHz is selected instead of $f_e$, because the phase difference between the currents conducting at the two coils is much larger in $f_e$, leading to the cancellation of the **E** field between the coils.

To summarize the choice of the frequency points for tri-mode operation, $f_2$ and $f_3$ are chosen based on equations (3) and (4), while $f_1$ is chosen based on the domination of current at coil 1 and the frequency being not too far away from $f_0$ (at around 5 % of $f_0$).

*D. Multi-Impedance Matching Network*

According to Fig. 4, the input impedance under different operating frequencies is different. In order to maximize the current conducting in the coils when they are driven by a 50 Ω pulse power amplifier, the input impedance of the three operation frequencies should be matched to 50 Ω with reactive components. Consider the three operation frequencies at 452.6 kHz, 457.7 kHz, and 480 kHz, their input impedances are 4.8 Ω, 24.1 Ω, and 2.6 + j 27 Ω, respectively. A multi-impedance matching network should be designed to match the three operation frequencies to 50 Ω effectively.

The multi-impedance matching network is shown in Fig. 6, which is integrated with the coil circuit. It is composed of a multi-section LC ladder network [27]. The values of the components are optimized by the gradient-based optimizer in Advanced Design System ® (ADS) [28] to match the operating frequencies to 50 Ω. The target of the matching is to make the scattering parameter $S_{11}$ to be smaller than -6 dB so that the reflected power is less than 25%, which is a practice or standard of impedance matching applied in mobile communication [23]. The optimization results $L_1$ = 11 μH, $L_2$ = 4.95 μH, $C_1$ = 33 nF, $C_2 = 56$ nF, and $C_3 = 56$ nF. From Fig. 5, it is shown that the current ratio between the coils does not vary with the cascaded matching network, so the impedance matching does not affect the tri-mode operation.

The matching network is implemented using lumped capacitors and inductors with soldering as shown in Fig. 7. It should be noticed that at lower frequency, the capacitors and inductors are difficult to be implemented using transmission line network since the wavelength is very large (666.7 m at 450 kHz), instead using lumped component is the best approach. The simulated and the measured scattering parameter $S_{11}$ are shown in Fig. 8. The discrepancy between the simulation result and the measurement result could be caused by the difference between the theoretical values and the actual implemented values of the lumped capacitors and inductors, and also the extra inductance due to the lead wires and soldering. To demonstrate the sensitivity of the circuit value, $L_1$ is changed to 10.4 $\mu$H in the simulation and the result is included in Fig. 7. As $L_1$ decreased, the lower band is spread to a wider band, while the higher band is shifted to a higher frequency. However, despite the discrepancy between the simulation and the actual circuit values, the measured $S_{11}$ of the three operation modes ($I_1>I_2$, $I_2>I_1$, $I_1=I_2$) has a return loss of smaller than -6 dB. The measurement result of the current ratio of the operation modes will be shown in Section V.

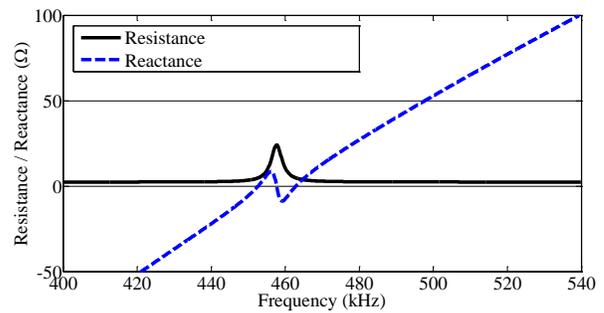

Fig. 4. Input impedance of the coil circuit.

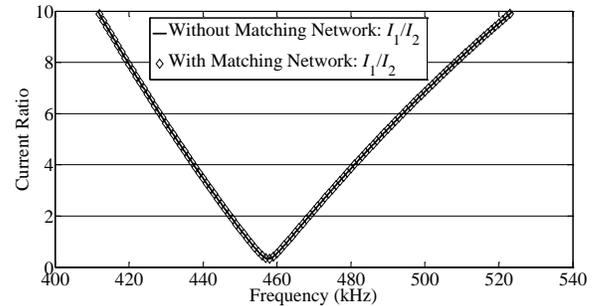

Fig. 5. The current ratio between the coils.

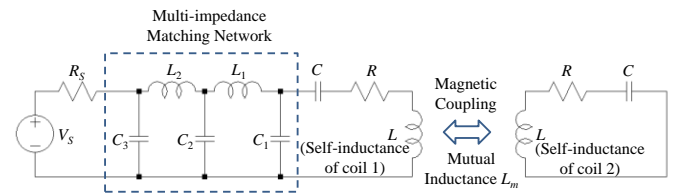

Fig. 6. The coil circuit integrated with the multi-impedance matching network.

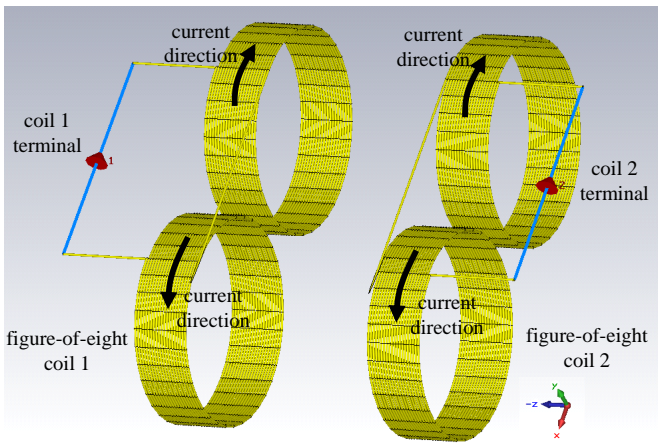

Fig. 2. A pair of magnetic coupled figure-of-eight coils.

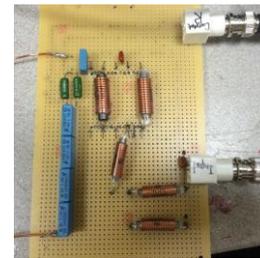

Fig. 7. Fabricated lumped components multi-impedance matching network.

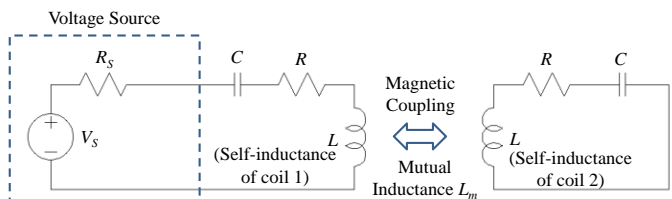

Fig. 3. The circuit configuration of the coil circuit.

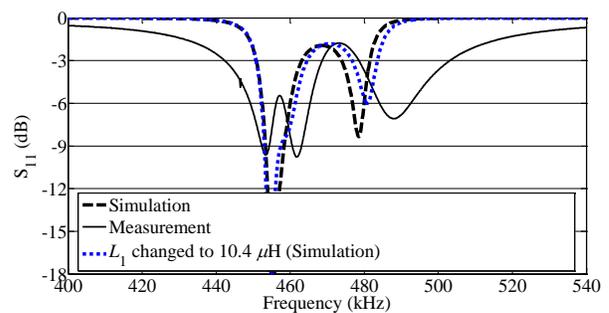

Fig. 8. The simulated and the measured return loss of the multi-impedance matched network.

## III. Electric Field Distribution of Coil

In this section, the electric field distribution of the coil under the three operation modes will be demonstrated.

### A. Mathematical Formula

Firstly, consider a circular coil of $N$ turns with negligible thickness, the electric field **E** generated by it can be computed by integrating along the whole current path C of the coil [22]:

$$\mathbf{E}(\mathbf{r}) = \frac{\mu_0 \omega N I}{4\pi} \oint_C \frac{\mathbf{dl'}}{|\mathbf{r} - \mathbf{r'}|}, \qquad (6)$$

where $\mu_0$ is the permeability, $N$ is the number of turn of the coil, $I$ is the magnitude of current flowing through the coil, $\mathbf{dl'}$ is the differential coil element, $\mathbf{r}$ and $\mathbf{r'}$ are the position vectors of the observation point and that of the differential element $\mathbf{dl'}$. Since the factor of $N$ is already in the equation, there is no need to integrate the path for each turn. Meanwhile, there are two coils (coils 1 and 2) as depicted in Fig. 2, the total **E** field at any observation point will be the superposition of the **E** field contributed by the two coils.

### B. Electric Field Distribution of the Tri-Mode Coil

As the theme of this paper, the electric field distribution of the coil in different operation mode is different. This is achieved by the different current ratio in the operation modes.

(i) First operation mode: $I_1 > I_2$

Consider the first operation mode where $I_1 > I_2$. From Fig. 5, $I_1/I_2 = 3.85$ at $f_1 = 480$ kHz, the calculated electric field distribution for every 1 A conducting coil 1 is shown in Fig. 9, where the phase difference between the two coils (which is not adjustable) is also taken into account. With the direction of the coordinate axis defined in Fig. 2, the origin of the $xyz$ coordinates is selected as the central point of the tri-mode coil setup, i.e. the central point in between coils 1 and 2. In the figure, coils 1 and 2 are located at $z = -4.5$ cm and 4.5 cm, respectively. It can be noticed that the **E** field is much stronger near coil 1. Furthermore, the 2D **E** and **B** field distribution on a contour color map obtained by electromagnetic stimulation (using CST Studio Suite ® [30]) at the plane $yz$-plane ($x = 0$) are shown in Fig. 10 and Fig. 11, respectively. It is observed that both the **E** and **B** field are much stronger near coil 1 than that of coil 2. Although there is some field leakage outside the coils, those regions are not within our interest of biomedical applications and hence can be ignored.

(ii) Second operation mode: $I_2 > I_1$

Consider the second operation mode where $I_2 > I_1$. From Fig. 4, $I_1/I_2 = 0.333$ at $f_2 = 457.6$ kHz, the calculated electric field distribution for every 1 A conducting coil 2 is also shown in Fig. 9. It can be noticed that the **E** field is much stronger near coil 2. Also, the 2D **E** and **B** field distribution on a contour color map obtained by electromagnetic stimulation at the plane $yz$-plane ($x = 0$) are shown in Fig. 12 and Fig. 13, respectively. It is observed that both the **E** and **B** field are much stronger near coil 2 than that of coil 1.

(iii) Third operation mode: $I_1 = I_2$

Consider the third operation mode where $I_1 = I_2$. From Fig. 4, $I_1 = I_2$ at $f_3 = 452.6$ kHz, the calculated electric field distribution for every 1 A conducting both coils is shown in Fig. 9. It can be noticed that the **E** field is strong at the both side of the coils and also superposed and enlarged the **E** field at the center. When comparing the three modes for **E** field distribution near the center, it is obvious that using the mode 3 is the most effective when the target is near the center. Also, the 2D **E** and **B** field distribution on a contour color map obtained by electromagnetic stimulation at the plane $yz$-plane ($x = 0$) are shown in Fig. 14 and Fig. 15, respectively. It is observed that both the **E** and **B** field are strong in between the both coils and at the center.

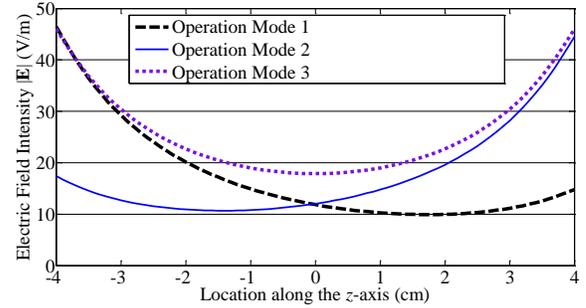

Fig. 9. The simulated **E** field distribution for the three operation modes

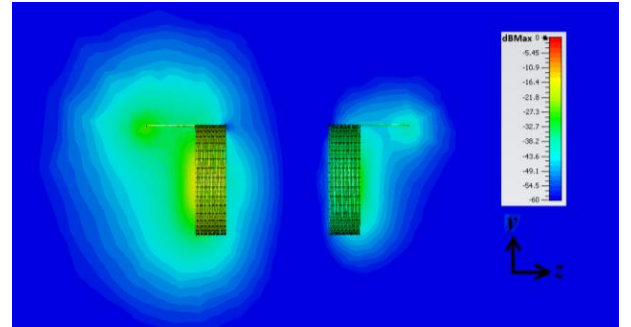

Fig. 10. The contour color map of **E** field distribution of the tri-mode coupled coil in the $yz$-plane ($x = 0$) in the first operation mode at $f_1 = 480$ kHz.

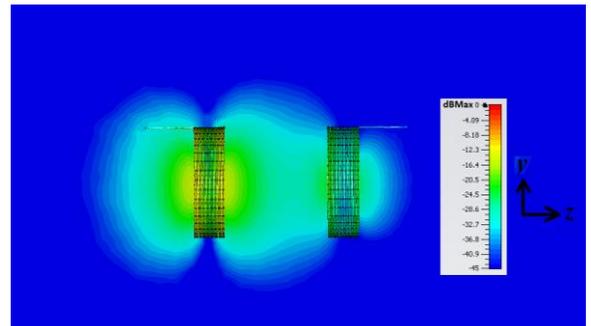

Fig. 11. The contour color map of **B** field distribution of the tri-mode coupled coil in the $yz$-plane ($x = 0$) in the first operation mode at $f_1 = 480$ kHz.

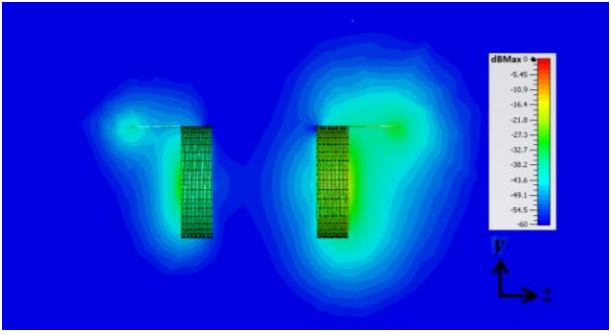

Fig. 12. The contour color map of **E** field distribution of the tri-mode coupled coil in the *yz*-plane ($x = 0$) in the second operation mode at $f_2 = 457.6$ kHz.

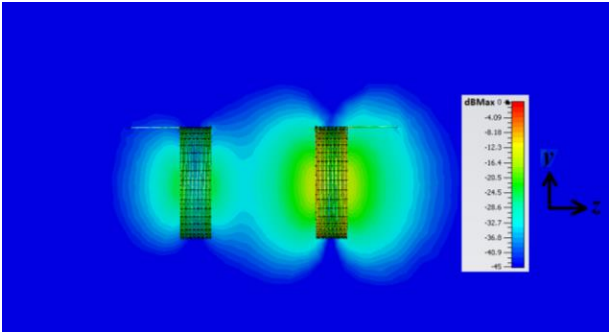

Fig. 13. The contour color map of **B** field distribution of the tri-mode coupled coil in the *yz*-plane ($x = 0$) in the second operation mode at $f_2 = 457.6$ kHz.

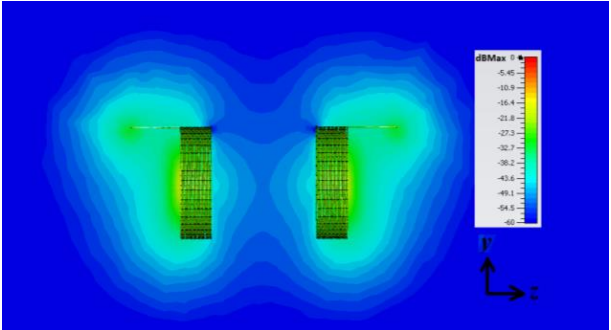

Fig. 14. The contour color map of **E** field distribution of the tri-mode coupled coil in the *yz*-plane ($x = 0$) in the third operation mode at $f_3 = 452.6$ kHz.

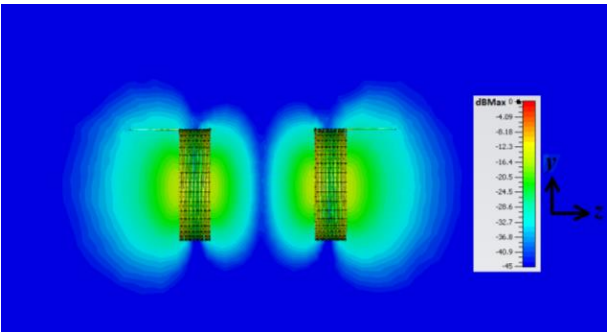

Fig. 15. The contour color map of **B** field distribution of the tri-mode coupled coil in the *yz*-plane ($x = 0$) in the third operation mode at $f_3 = 452.6$ kHz.

## IV. Focus Tuning Procedures

In this section, the focus tuning procedures when using the coils will be discussed.

### A. Operation Mode Selection

As depicted in the previous sections, when the excitation location is near coil 1, the operation mode 1 is preferred which would weight low the current conducting at coil 2. It is because coil 2 is not effective to induce **E** field at the excitation point, but it would increase the **E** field near coil 2.

In contrast, when the excitation location is near coil 2, the operation mode 2 is preferred which would weight low the current conducting at coil 1. This would minimize the **E** field near coil 1 which is not near the excitation point.

When the excitation location is near the center, operation mode 3 is preferred which would conduct current equally at both coils because the E field generated by both coil is effective to increase the **E** field level at the excitation point. Since the **E** field attenuates with distance from the source, using only one coil would need a much larger current to generate adequate **E** field magnitude at the center, compared with the locations near the coils. Hence, the mode 3 which utilizes both the current at both coils would be effective to prevent overloading of current at the either coil and the **E** field magnitude near it, by superposing the **E** field from both coils at the center. Therefore, depending on the *z*-coordinates of the excitation location, the operation modes are selected as follows:

(i) Mode 1: $\dfrac{-D}{2} < z < \dfrac{-D}{6}$

(ii) Mode 2: $\dfrac{D}{6} < z < \dfrac{D}{2}$

(iii) Mode 3: $\dfrac{-D}{6} \leq z \leq \dfrac{D}{6}$

### B. Voltage Level Determination

In the determination of the voltage levels at the sources, the ratio between the source voltage and the current conducting at the coils, as well as the ratio between the electric field and the current conducting at the coils, should be calculated.

Firstly, from circuit theory, the current conducting at the coil is proportional to the source voltage. The transconductance between the coil currents ($I_1$ and $I_2$) and the source voltage $V_S$ are given by:

$$G_{I1} = \frac{I_1}{V_S} \quad (7)$$

$$G_{I2} = \frac{I_2}{V_S} \quad (8)$$

Secondly, from electromagnetic theory [18], the electric field induced at the observation point is proportional to the coil current as well as the frequency. The electric field gain between the electric field induced at the observation point and the product of the coil current and frequency is given by:

$$G_{E1} = \frac{E_1}{I_1 f}, \quad (9)$$

$$G_{E2} = \frac{E_2}{I_2 f}, \quad (10)$$

where $E_1$ and $E_2$ are the magnitudes of the electric field intensity induced at the observation point contributed by coil 1 current $I_1$ and coil 2 current $I_2$, respectively. Then, from the above equations, the electric field induced at the observation point can be related to the source voltage by:

$$E_1 = G_{E1} G_{I1} f V_S, \quad (11)$$
$$E_2 = G_{E2} G_{I2} f V_S. \quad (12)$$

The overall voltage induced by the two coils is then the superposition of the **E** field from the above formula, given by:

$$E = \sqrt{(G_{E1} G_{I1})^2 + (G_{E2} G_{I2})^2 + 2 G_{E1} G_{I1} G_{E2} G_{I2} \cos\theta} \cdot f V_S, \quad (13)$$

where $\theta$ is the phase difference between the currents $I_1$ and $I_2$, respectively. Therefore, the voltage of the source $V_S$ can be calculated from the required **E** field at the observation point:

$$V_S = \frac{E}{\sqrt{(G_{E1} G_{I1})^2 + (G_{E2} G_{I2})^2 + 2 G_{E1} G_{I1} G_{E2} G_{I2} \cos\theta} \cdot f}. \quad (14)$$

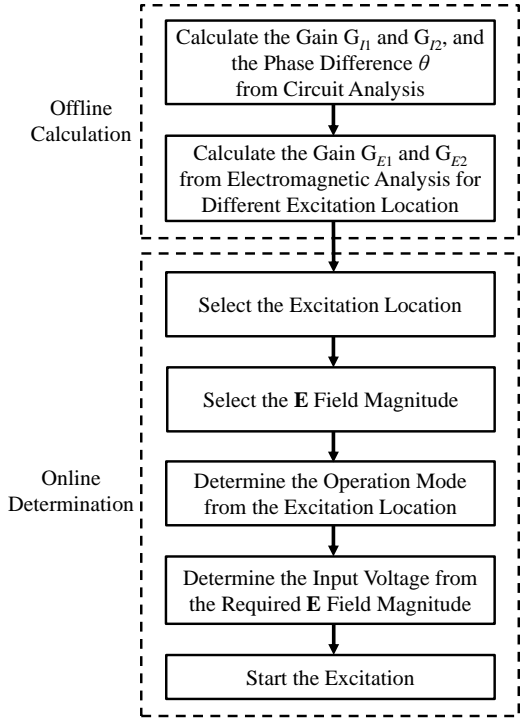

Fig. 16. The flowchart of selecting the operation mode and the source input voltage.

*C. Overall Flowchart*

The procedure to determine which mode to be chosen as well as the source voltage is given in Fig. 16. Before the start of the excitation of the coil, the gains $G_{I1}$, $G_{I2}$, $G_{E1}$ and $G_{E2}$, and the phase difference $\theta$ should be calculated from the circuit analysis and electromagnetic analysis. The gains $G_{E1}$ and $G_{E2}$ are dependent on the excitation locations where the corresponding values at different locations can be calculated and stored in a database. After the calculation, the user can determine the excitation location and the **E** field magnitude according to the application. For example, in magnetic stimulation, the magnitude of the **E** field is selected as 150 V/m in the literature [31], [32]. Then, the operation mode is calculated by the excitation location, and the source input voltage is calculated by the required **E** field magnitude. After all, the coil circuit can be excited to generate a required **E** field level at the excitation location while minimizing the **E** field level out of the target.

*D. Verification of Formulas*

In this subsection, the focus tuning formulas would be applied at different excitation locations for verifications. Three typical excitation locations would be demonstrated, which are $z = -2$ cm, $z = 0$ cm, and $z = 3$ cm, where each of them corresponds to each of the three operation modes, respectively. The **E** field magnitude for excitation is set as 150 V/m, as previous applied for magnetic stimulation in the literature [31], [32]. For the procedure of operation, the flowchart as given in Fig. 16 can be referred.

(i) Excitation at $z = -2$ cm with 150 V/m

When the excitation location is at $z = -2$ cm, operation mode 1 would be applied. From the circuit analysis in Fig. 6, $G_{I1}$ equals $I_1/V_S = 0.038$ A/V and $G_{I2}$ equals $I_2/V_S = 0.010$ A/V with the phase difference $\theta = 175.3°$, at operation mode 1 using $f_1 = 480$ kHz. From Eqts. (9) and (10), $G_{E1} = 3.941 \times 10^{-5}$ (V/AmHz$^{-1}$) and $G_{E2} = 1.124 \times 10^{-5}$ (V/AmHz$^{-1}$). According to Eqt. (14), the calculated source voltage $V_S = 225.5$ V, so that the coil currents are driven as $I_1 = 8.57$ A and $I_2 = 2.26$ A. The calculated electric field distribution is shown in Fig. 17, which verified that the **E** field intensity at $z = -2$ cm is 150 V/m.

(ii) Excitation at $z = 3$ cm with 150 V/m

When the excitation located is at $z = 3$ cm, operation mode 2 would be applied. From the circuit analysis in Fig. 6, $G_{I1}$ equals $I_1/V_S = 0.014$ A/V and $G_{I2}$ equals $I_2/V_S = 0.041$ A/V with the phase difference $\theta = 88.1°$, at operation mode 2 using $f_2 = 457.6$ kHz. From Eqts. (9) and (10), $G_{E1} = 8.647 \times 10^{-6}$ (V/AmHz$^{-1}$) and $G_{E2} = 5.898 \times 10^{-5}$ (V/AmHz$^{-1}$). According to Eqt. (14), the calculated source voltage $V_S = 135.1$ V, so that the coil currents are driven as $I_1 = 1.89$ A and $I_2 = 5.54$ A. The calculated electric field distribution is also shown in Fig. 17, which verified that the **E** field intensity at $z = 3$ cm is 150 V/m.

(iii) Excitation at $z = 0$ cm with 150 V/m

When the excitation located is at $z = 0$ cm, operation mode 3 would be applied. From the circuit analysis in Fig. 6, $G_{I1}$ equals $I_1/V_S = 0.029$ A/V and $G_{I2}$ equals $I_2/V_S = 0.028$ A/V with the phase difference $\theta = 19.5°$, at operation mode 3 using $f_3 = 452.6$ kHz. From Eqts. (9) and (10), $G_{E1} = G_{E2} = 2.004 \times 10^{-5}$ (V/AmHz$^{-1}$). According to Eqt. (14), the calculated source voltage $V_S = 294.4$ V, so that the coil currents are driven as $I_1 = 8.54$ A and $I_2 = 8.24$ A. $V_S$, $I_1$ and $I_2$ are the largest within the three cases because the distance from the nearest coil is farther. The calculated electric field distribution is also shown in Fig. 17, which verified that the **E** field intensity at $z = 0$ cm is 150 V/m.

In general, based on the operation mode selection and the desired **E** field level of 150 V/m, the calculated source voltage for different excitation locations using equation (14) are given in Fig. 18. This verified that the excitation location can be adjusted to various locations between the two coils by selecting the operation mode and the source voltage $V_s$. The figure indicates that the farther the distance from the coils, the larger the source voltage required for generation of similar **E** field magnitude at the excitation location. This is due to the fact that **E** field attenuates with distance from the coils, and therefore a larger source voltage would be required to conduct larger currents to the coils and hence generate an adequately large magnitude of **E** field at a farther excitation location. For example, the excitation location of $z = -2$ cm requires $V_s = 207.0$ V, while the excitation location of $z = -2.2$ cm requires $V_s = 225.5$ V, when both are calculated for 150 V/m.

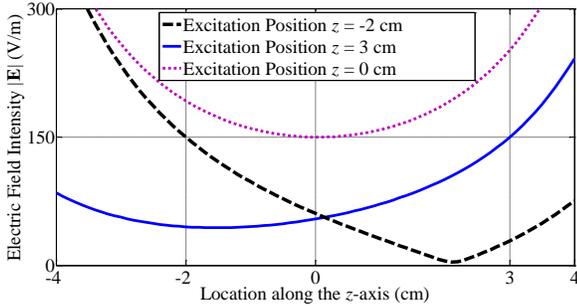
Fig. 17. The simulated **E** field distribution for the three excitation positions to achieve an intensity of 150 V/m.

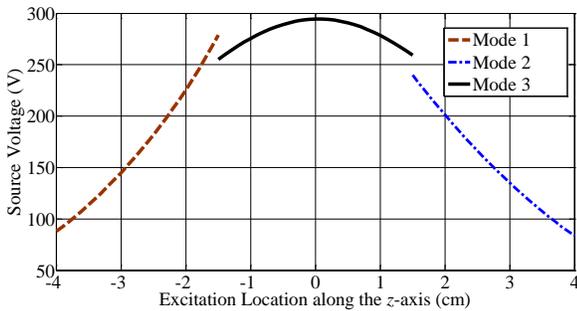
Fig. 18. The calculated source voltage $V_s$ for different excitation location in the three operation modes to achieve an intensity of 150V/m.

## V. COIL MEASUREMENT

In this section, the fabricated tri-mode coupled coil as shown in Fig. 19 is measured to verify the calculation results. In the measurement, the tri-mode coupled coil is driven using a 50 Ω pulse amplifier with signal generator. The currents flowing through the coils are measured by detecting the voltage across the small resistor in series with the coil (as mentioned in Section II). Since there exists differences between the theoretical values and the actual implemented circuit component values, the actual frequencies of the three operation modes are little bit shifted from the simulated operation frequencies.

### A. First Operation Mode: $I_1 > I_2$

Consider the first operation mode where $I_1 > I_2$. From Fig. 8, the measured $S_{11}$ is of local minimum at 487.7 kHz. Since this frequency point is away from the resonance frequency to give $I_1 > I_2$ with impedance matching done to minimize the return loss, it is selected as the first operation mode as $f_1 = 487.7$ kHz.

At this frequency, the MRC coil is driven with a pulse of 1 ms duration so that the voltage is controlled to give coil 1 current of around 5 A, where the currents of both coils are recorded in Fig. 20. It should be noticed that current has a short transient response of around 0.2 ms before it reaches the steady state.

At the steady state, the measured current ratio of $I_1/I_2 = 13.0$. According to this measured ratio, the calculated electric field distribution for every 1 A conducting at coil 1 according to Eq. (6) is shown in Fig. 21. Again, coil 1 and coil 2 are located at $z = -4.5$ cm and 4.5 cm, respectively. It can be noticed that the **E** field is much stronger near coil 1.

### B. Second Operation Mode: $I_2 > I_1$

Consider the second operation mode where $I_2 > I_1$. From the measurement, the frequency which results a minimum ratio of $I_1/I_2$ is measured and used as the second operation mode, which is $f_2 = 453.2$ kHz.

At this frequency, the MRC coil is driven with a pulse of 1 ms duration so that the voltage is controlled to give coil 2 current of around 5 A, where the currents of both coils are recorded in Fig. 22. Again, it should be noticed that current has a short transient response of around 0.2 ms before it reaches the steady state.

At the steady state, the measured current ratio of $I_1/I_2 = 0.258$. According to this measured ratio, the calculated electric field distribution for every 1 A conducting coil-1 according to Eq. (6) is also shown in Fig. 21. It can be noticed that the **E** field is much stronger near coil 2.

### C. Third Operation Mode: $I_1 = I_2$

Consider the third operation mode where $I_1 = I_2$. From the measurement, the frequency which result $I_1 = I_2$ below $f_2$ is measured and used as the third operation mode, which is $f_3 = 451.1$ kHz.

At this frequency, the MRC coil is driven with a pulse of 1 ms duration so that the voltage is controlled to give the current across both coils of around 5 A, where the currents of both coils are recorded in Fig. 23. Again, it should be noticed that current has a short transient response of around 0.2 ms before it reaches the steady state.

At the steady state, the measured current ratio of $I_1/I_2 = 1.03$. According to this measured ratio, the calculated electric field distribution for every 1 A conducting at coil 1 according to Eq. (6) is also shown in Fig. 21. Again, it can be noticed that the **E** field is strong at the both side of the coils and also superposed and enlarged the **E** field at the center.

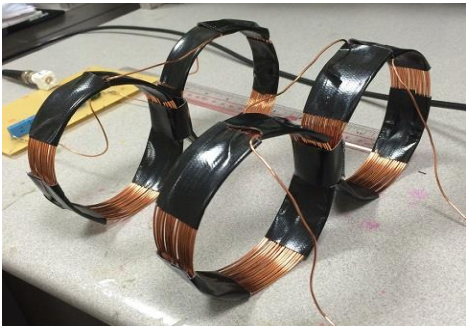

Fig. 19. The fabricated coupled coil for measurement.

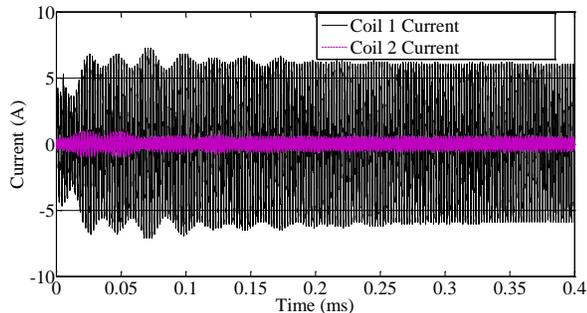

Fig. 20. The measured current ratio is given for the operation mode 1 at $f_1$ = 487.7 kHz.

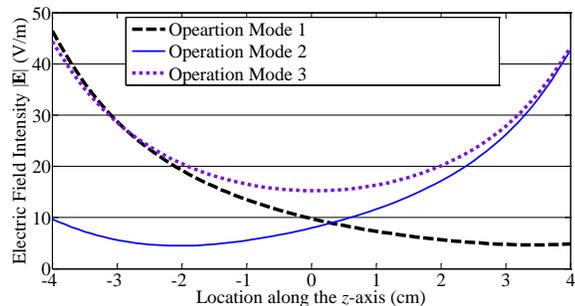

Fig. 21. The calculated **E** field distribution according to the measured current ratio.

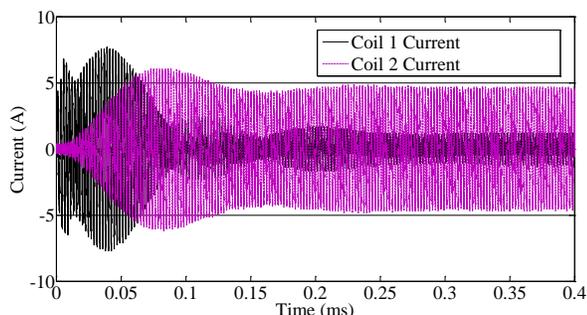

Fig. 22. The measured current ratio is given for the operation mode 2 at $f_2$ = 453.2 kHz.

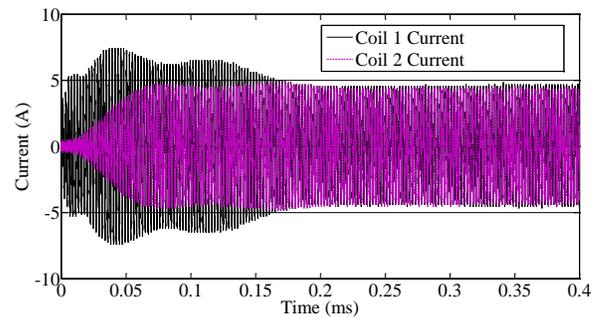

Fig. 23. The measured current ratio is given for the operation mode 3 at $f_3$ = 451.1 kHz.

## VI. Ex-vivo Tissue Measurement

In this section, ex-vivo tissue experiment [22] will be done to verify the tri-mode concept on field distribution. In the experiment, the tri-mode coupled coil is driven using a 50 Ω pulse amplifier with signal generator on the three operation modes. Then, a piece of pig muscle tissue of the size 1.0 cm × 2.0 cm × 3.5 cm is placed in between the coils to induce voltage and current on it through electromagnetic induction, as illustrated in Fig. 24. Electrodes are used to measure the induced voltage across it. For each of the operation mode, the induced voltage at three positions (i) near coil 1: $z$ = -2 cm, (ii) at the center: $z$ = 0 cm, (iii) near coil 2: $z$ = 2 cm, are compared by relocating the position of the tissue. This can verify the tri-mode concept on the field distribution.

First of all, the signal generator voltage is fixed at 0.5 $V_{peak}$ with the frequency 451.1 kHz, then the source voltage from the pulsed amplifier becomes $Vs$ = 451.5 $V_{peak}$, which induced voltage across the tissue under mode 3 (451.1 kHz) where both coils 1 and 2 are delivering magnetic field with the same current. The induced voltages at the three positions $z$ = -2, 0, and 2 cm are shown in Figs. 25 – 27, respectively. From the figures, although the voltage is contributed by the magnetic field radiated from the two coils, the transient response when the tissue is placed at different locations are similar and the steady state is as short as around 0.1 ms in all the three cases.

Next, the voltage measurement of the tissue is repeated at the mode 1 (487.7 kHz) and mode 2 (453.2 kHz) to find the relative magnitude of the voltage induced at the positions $z$ = -2, 0, and 2 cm. The magnitude of the voltage measured is then normalized to the maximum and compared in Table I. The data shown in the table completely verified the concept of the tri-mode coil: In the mode 1, the voltage induced near coil 1 would be strongest while being weak near coil 2. In mode 2, the voltage induced near coil 2 would be strongest while being weak near coil 1. In mode 3, the fact that the voltage induced is strongest near the coils can also be found in the experiment result, where the voltage induced would be decreased with the increase of the distance from the coils. However, the electromagnetic induction is contributed by both coils and hence decay fewer along the distance among the three modes, thereby resulting a larger induced voltage at the center.

Despite of the successful experiment results, it should be pointed out that an assumption of the experiment is the constant resistivity of the tissue within the operation.

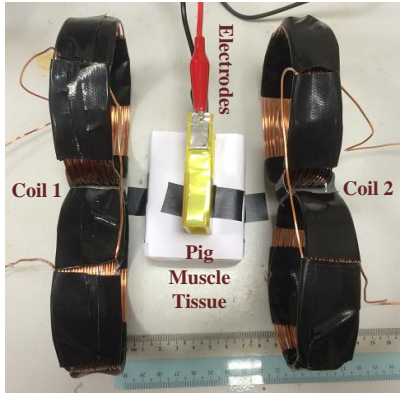

Fig. 24. Illustration of the ex-vivo experiment.

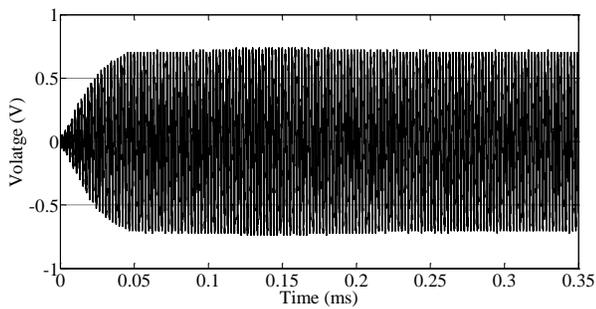

Fig. 25. The induced voltage at the tissue at $z = -2$ cm (tissue placed at the center).

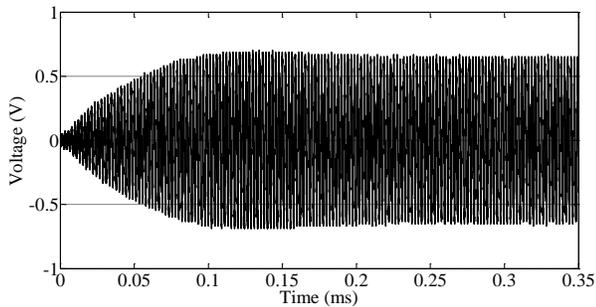

Fig. 26. The induced voltage at the tissue at $z = 0$ cm (tissue placed near coil 1).

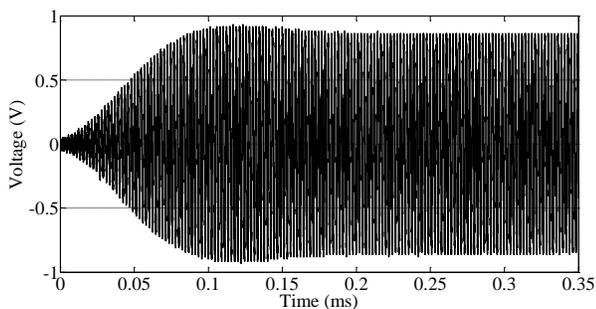

Fig. 27. The induced voltage at the tissue at $z = 2$ cm (tissue placed near coil 2).

TABLE I
THE NORMALIZED MEASURED VOLTAGE AT DIFFERENT LOCATIONS AND DIFFERENT OPERATING MODES

|  | $z = -2$ cm (The tissue is placed near the coil 1) | $z = 0$ cm (The tissue is placed at the center) | $z = 2$ cm (The tissue is placed near the coil 2) |
|---|---|---|---|
| Mode 1 (Current is strong at coil 1) | 1.000 | 0.524 | 0.286 |
| Mode 2 (Current is strong at coil 2) | 0.376 | 0.464 | 1.000 |
| Mode 3 (Current is strong at both coil) | 0.816 | 0.755 | 1.000 |

VII. DISCUSSION ON PRACTICAL ISSUES

In this section, several practical issues of the tri-mode coil design will be discussed for further clarification of the design.

*A. Comparing different coil configuration*

In this subsection, the use of a pair figure-of-eight coils are compared to a pair of circular coils and square coils. The calculated **E** field distribution of the compared coils along the *x*- and *y* –axis at *z* = 0 cm for every 1 A conducting at both coils 1 and 2 is shown in Figs. 28 and 29. This is the situation in mode 3, where both coils are conducting the same current. The origin of the coordinates is selected as the central point of the coupled coils. For ease of analysis in the comparison, the phase difference between coils 1 and 2 is assume to be zero of that the **E** field contributed from the two coils are added up totally. For a fair comparison, the coils are compared in the same number of turns. The diameters of the figure-of-eight coils and the circular coil, as well as the diagonal of the square coil are all 9 cm. From the comparison, it is obvious that the main benefit of using a pair of figure-of-eight coil is the finer focusing of **E** field along the *x*- and the *y*- axis at *x* = *y* = 0 cm.

*B. The symmetry of the E field in the mode 3*

From the literature [33], [34], it is found that there are some secondary effects on the generated magnetic field and consequently the generated electric field due to manufacturing mismatches or assembly misalignments in the circular coils.

It can be concluded easily by inspection that there might be the same effects or even more another experimental imperfection effects in a pair of 8-shape coils, too. Therefore, by considering the given simple prototype system (refer to Fig. 19), it might be difficult to achieve symmetrical experimental E fields distributions as given in Fig. 21 (Operation Mode 3).

For the above concerns, it is necessary to observe how large the error is introduced to affect the symmetry of **E** field. This can be observed by the experiment in Section VI, where the voltage induced at the tissue is measured at different location along the z-axis. There is a relation between the voltage *V* and the **E** field induced [35]

$$V = d \, |\mathbf{E}| \qquad (15)$$

where *d* is the distance between the two terminals of the voltage measurement, and the assumption is constant |**E**| magnitude between the two terminals. Therefore, according to this equation, the tissue measurement in Section VI can be used as an **E** field probe to estimate the field induced at the tissue.

Another trial of the tissue measurement is done in operation mode 3 with more locations involved from $z = -3$ cm to $z = 3$ cm, while the result is presented in Fig. 30. Practically, it can be observed that the voltages measured at the two sides are quite symmetric (with maximum imperfection 12.5 % at $z = 2$ cm), although they are not perfectly equal to each other.

*C. Effect of the medium in between the coils*

It should be noticed that equation (6) is used to calculate the **E** field distribution. However, this **E** field formulation is valid only in free (homogeneous) space. If there is a conductive body between the coils, the **E** field distribution will significantly change from that given by equation (6) due to different medium is involved. For example, although the pork tissue used in the previous section is a poor conductor ($\sigma = 0.8$ S/m [18]), it still have large effect on the **E** field induced. This can be observed by the following time harmonic equation derived from the ampere law [18], [35]:

$$\mathbf{E} = \frac{1}{\sigma + j\omega\varepsilon}\nabla\times\mathbf{H} \qquad (16)$$

It is clear that the **E** field induced by electromagnetic induction must be much smaller in the tissue, where the change of magnitude in different medium is reflected by the denominator term.

However, the accurate information of biological medium might not be included during the design of the coils, because the biological medium to be placed in between the models varies case by case. For instance, considering different patients have different body size, using a fixed model of biological medium would be impractical.

Instead, the free space condition could be used as a reference in the design process. This is because a 450 kHz magnetic flux should be able to penetrate the tissue which is a poor conductor with high dielectric constant (some previous simulation from the author group in a MRC coil design could be found at [22] and [35]). The penetration of magnetic flux is similar to that of the traditional magnetic stimulation [1], [6] – [8] which operates at lower frequencies. Then, from equation (16), the penetrated magnetic flux should be able to induce **E** field at the tissue. If a larger magnitude of **H** is involved in equation (6) (e.g. due to a larger magnitude of current passing through the coils), the larger should be the magnitude of the induced **E** no matter the medium is tissue or air. This should explain why free space is used as a design reference in this paper. For a comparison with the literature [8], the shape of **E** field distribution as observed in Fig. 28 for a figure-of-eight coil simulation under free space is similar to that of the slinky-2 coil (i.e. figure-of-eight coil) from the experimental measurement under saline solution in [8, Fig. 3].

For further verification, the simulated **E** field distribution of mode 3 with a piece of muscle tissue ($\sigma = 0.8$ S/m, $\varepsilon_r = 56.8$ [22]) of the size 24 cm × 12 cm × 4 cm is shown in Fig. 31. It is verified that the **E** field is much smaller in the tissue than in the air, while there is still a focusing pattern inside the tissue. The simulated field distribution is a good agreement with that calculated in Figs. 9 and 28.

*D. Recapping the objective of the tri-mode operation*

From literature study, it might not be possible to use any combination of coils placed outside the conductor to focus the **E** field in depth in spherical geometry [36]. The largest **E** field always occurs at the boundary of such a conductor. Moreover, from Figs. 21 and 30, it is clear that the magnitude of **E** field attenuates with the increase of distance between the coils and the observation point. This does not imply that the design is not valid. Actually, the objective of the tri-mode design is not to create a local maximum of **E** field along the *z*-axis as indicated in Figs. 21 and 30.

Instead, the objective of the tri-mode coil design is to generate three different operations modes so as to make a better electromagnetic field distribution in different operation scenario. To recap the statements in Section II C: Mode 1 is used to minimize the electromagnetic field distribution near coil 2 where is not near the focal point located near coil 1; Mode 2 is used to minimize the electromagnetic field distribution near coil 1 where is not near the focal point located near coil 2. Mode 3 is used to prevent overloading the current at the either coil and the **E** field magnitude nearby by utilizing both coils to superpose a large **E** field at the center.

*E. Bio-medical Applications*

Physically, the coil design provides two functions: (i) inducing current at biological medium by electromagnetic induction, which can be observed by the equations (1) and (2); (ii) heating the biological medium from the electric power produced by the induced current, where the ohmic power density $P_\sigma$ under the conduction of current is given by

$$P_\sigma = \frac{|\mathbf{J}|^2}{\sigma} = \sigma|\mathbf{E}|^2 \qquad (17)$$

The current induced and hence the heating generated at the biological medium could be used for biological applications. For instance, the pulsed electromagnetic energy treatment [8] considers the delivery of pulsed RF energy to a tissue for therapeutic effects. The coils in this paper are designed at 450 kHz that is in the mid frequency range of the pulsed electromagnetic energy treatment [9, Table 3]. Furthermore, the thermoacoustic imaging [12] applies heating for thermal expansion of tissue which leads to the generation of an acoustic wave for imaging purposes. Since the above depends on current induced in the tissue by electromagnetic induction, the current distribution at a piece of muscle tissue is simulated (at mode 3) and shown in Fig. 32. Since air is not conductive, there is no current induced at the air region although there is **E** field. Instead, current is induced along the piece of tissue. The current induced in the tissue would cause the generation of heat due to equation (17), which could potentially be applied for pulsed electromagnetic energy treatment [9] and thermoacoustic imaging [12] applications.

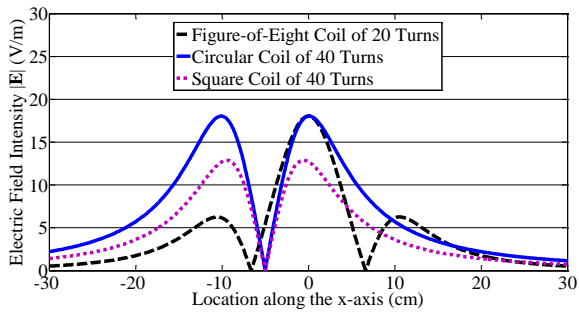

Fig. 28. **E** field distribution of different coils along the *x*-axis.

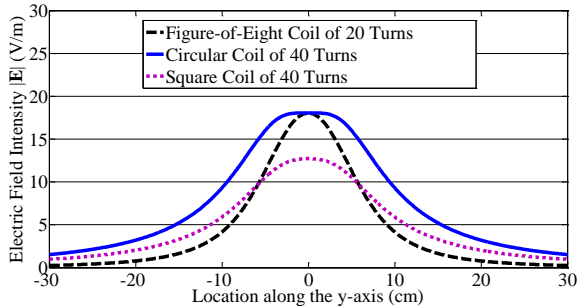

Fig. 29. **E** field distribution of different coils along the *y*-axis.

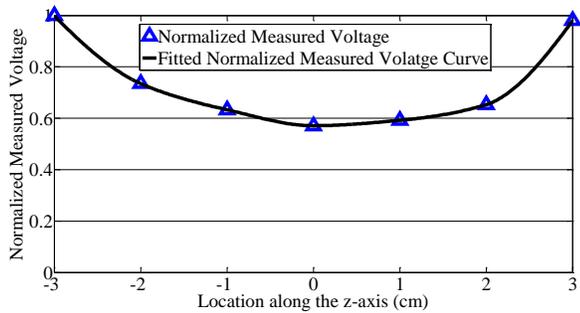

Fig. 30. The normalized measured voltage at different locations at mode 3.

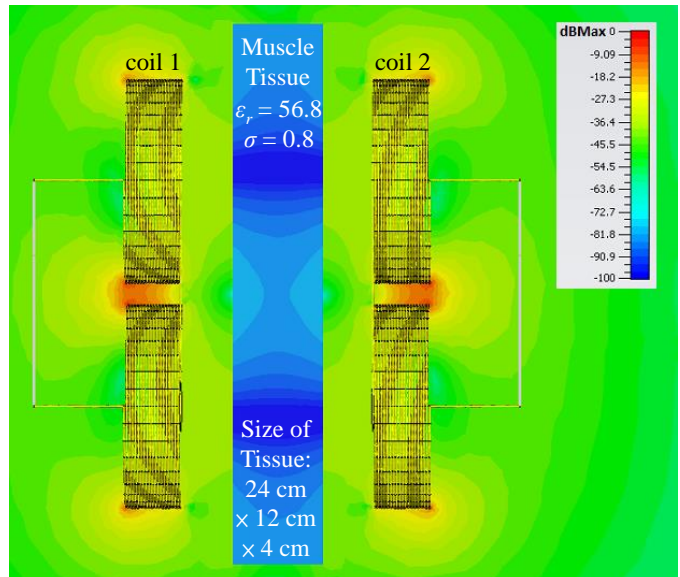

Fig. 31. E field distribution of mode 3 in the *xz*-plane ($y = 0$) with a muscle tissue of the size 24 cm × 12 cm × 4 cm.

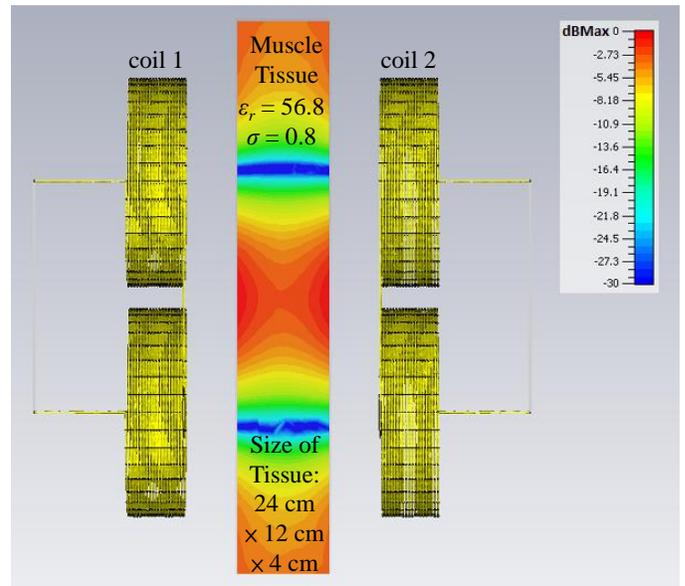

Fig. 32. Current distribution of mode 3 in the *xz*-plane ($y = 0$) with a muscle tissue of the size 24 cm × 12 cm × 4 cm.

## VIII. CONCLUSION

In this paper, a tri-mode coil has been designed which operates in three operation modes. The **E** field distribution can be configured by changing the current ratio in coil 1 and coil 2. A focus tuning procedure has been formulated which determines the operation mode and the source voltage to be applied according to the excitation location and the **E** field intensity at the target. The formulas have been verified through an example which could generate 150 V/m at the target location $z = -2$ cm, 0 cm, and 3 cm. The proposed method can be applied for different constrains in different applications. The concept of the tri-mode in electromagnetic induction is also verified in the ex-vivo tissue experiment in Section VII. The symmetry of the field distribution in the mode 3 has also been demonstrated and discussed. The limitation of the setup would be the fact that only three modes are useful for different field distribution with the current setup of two coils. Therefore, in future work, more coils or a coil array can be applied instead of using just coils 1 and 2. With more coils involved, the ratio between the current across the coils and thus the **E** field distribution might have more combinations, which could further optimize the induction of current in the biological medium as well as achieving multiple excitation targets. Moreover, the actual implementation of the biological applications such as pulsed electromagnetic energy treatment [9] and thermoacoustic imaging [12] are also suggested to be the future work.

## ACKNOWLEDGMENT

This research is supported by the Singapore National Research Foundation under Exploratory/Developmental Grant (NMRC/EDG/1061/2012) and administered by the Singapore Ministry of Health's National Medical Research Council.